\documentclass[a4paper,fleqn,usenatbib]{mnras}

\usepackage{mathptmx}

\usepackage[T1]{fontenc}
\usepackage{ae,aecompl}

\usepackage{graphicx}	
\usepackage{amsmath}	
\usepackage{amssymb}	
\usepackage{url}

\usepackage{array}
\usepackage{footnote}
\usepackage{rotating}
\usepackage{natbib}
\usepackage{multicol}
\usepackage{multirow}
\usepackage{lscape}
\usepackage[usenames, dvipsnames]{color}
\usepackage{longtable}
\usepackage{tablefootnote}

\def\oc{\textit{O$-$C}}
\def\lt{LiTE}
\def\bl{Blazhko}
\def\oc{\textit{O$-$C}}
\def\f0{$f_{0}$}
\def\fbl{$f_{\rm BL}$}
\def\fm{$f_{\rm M}$}
\def\kf0{$kf_{0}$}
\def\pbl{$P_{\rm BL}$}
\def\ppuls{$P_{\rm Puls}$}

\title[Z CVn - Blazhko star in binary?]{A cautionary tale of interpreting O-C diagrams: period instability in a classical RR Lyr Star Z CVn mimicking as a distant companion}

\author[Skarka et al.]{
\begin{minipage}{1.0\textwidth}
M. Skarka$^{1,3}$\thanks{marek.skarka@csfk.mta.hu}, 
J. Li{\v s}ka$^{2,3}$, 
R. D\v{r}ev\v{e}n\'{y}$^{3,4}$, 
E. Guggenberger$^{5,6}$, 
\'{A}. S\'{o}dor$^{1}$,
T. G. Barnes$^{7}$ 
and K. Kolenberg$^{8,9}$  \vspace*{0.3cm}
\end{minipage}\\ 
$^{1}$ Konkoly Observatory, Research Centre for Astronomy and Earth Sciences, Hungarian Academy of Sciences, \\ ~~Konkoly Thege Mikl\'{o}s \'{u}t 15-17, H-1121 Budapest, Hungary \\
$^{2}$ Central European Institute of Technology - Brno University of Technology (CEITEC BUT), Purky\v{n}ova 656/123, CZ-612 00 Brno, \\ ~~Czech Republic\\
$^{3}$ Variable Star and Exoplanet Section of the Czech Astronomical Society, Vset\'{i}nsk\'{a} 78, Vala\v{s}sk\'{e} Mezi\v{r}\'{i}\v{c}\'{i}, CZ-757 01, Czech Republic \\
$^{4}$ Znojmo Observatory, Vinohrady 57, CZ-669 02 Znojmo, Czech Republic \\
$^{5}$ Max Planck Institut f\"{u}r Sonnensystemforschung, Justus-von-Liebig-Weg 3, D-37077 G\"{o}ttingen, Germany \\
$^{6}$ Stellar Astrophysics Centre, Department of Physics and Astronomy, Aarhus University, DK-8000 Aarhus C, Denmark\\
$^{7}$ The University of Texas at Austin, McDonald Observatory, 1 University Station, C1402, Austin, TX 78712, USA\\
$^{8}$ Institute of Astronomy, KU Leuven, Celestijnenlaan 200D, B-3001 Heverlee, Belgium \\
$^{9}$ Physics Department, University of Antwerp, Groenenborgerlaan 171, B-2020 Antwerpen, Belgium
}

\date{Accepted October 17. Received October 16; in original form August 18}

\pubyear{2017}

\begin{document}
\label{firstpage}
\pagerange{\pageref{firstpage}--\pageref{lastpage}}
\maketitle

\begin{abstract}
We present a comprehensive study of Z CVn, an RR Lyrae star that shows long-term cyclic variations of its pulsation period. A possible explanation suggested from the shape of the \oc~diagram is the light travel-time effect, which we thoroughly examine. We used original photometric and spectroscopic measurements and investigated the period evolution using available maximum times spanning more than one century. If the binary hypothesis is valid, Z CVn orbits around a black hole with minimal mass of $56.5$\,$\mathfrak{M}_{\odot}$ on a very wide ($P_{\rm orbit}=78.3$\,years) and eccentric orbit ($e=0.63$). We discuss the probability of a formation of a black hole--RR Lyrae pair and, although we found it possible, there is no observational evidence of the black hole in the direction to Z CVn. However, the main objection against the binary hypothesis is the comparison of the systemic radial velocity curve model and spectroscopic observations that clearly show that Z CVn cannot be bound in such a binary. Therefore, the variations of pulsation period are likely intrinsic to the star. This finding represents a discovery/confirmation of a new type of cyclic period changes in RR Lyrae stars. By the analysis of our photometric data, we found that the \bl~modulation with period of 22.931\,d is strongly dominant in amplitude. The strength of the phase modulation varies and is currently almost undetectable. We also estimated photometric physical parameters of Z CVn and investigated their variations during the Blazhko cycle using the Inverse Baade-Wesselink method.

\end{abstract}

\begin{keywords}
Methods: data analysis -- techniques: spectroscopic -- techniques: photometric -- stars: variables: RR Lyrae -- stars: binaries: general -- stars: individual: Z~CVn
\end{keywords}

\section{Introduction}\label{Sect:Introduction}
A significant portion of stars in the Universe are known to be bound in double or multiple systems \citep[e.g.][]{abt1983,guinan2006}. Pulsating stars in such systems are of particular interest because their physical parameters can be determined independently both from pulsation characteristics and from investigation of their orbital motion \citep[][]{prada2012,lee2017}. Moreover, eclipsing binaries offer the only chance to directly determine the mass of both components with very good accuracy without any additional assumptions \citep{helminiak2009,kiefer2016}. 

Many pulsating stars of various types have been identified in binary systems \citep[see, for example, the lists by][]{szabados2003,soydugan2006,zhou2010}. However, stars of RR Lyrae (RRL) type seem to be an exception. Apart from TU UMa, which is quite probably in binary with low-mass companion \citep{wade1999, liska2016a}\footnote{Unambiguous spectroscopic confirmation is still missing even in TU UMa.}, not a single classical RRL has been unambiguously confirmed to reside in a binary system so far. There are only several tens known RRL binary candidates \citep[see the online version of the list by][]{liska2016b,liska2016c}\footnote{\url{http://rrlyrbincan.physics.muni.cz/}} out of more than one hundred thousands of catalogued RRL stars \citep[field RRLs plus RRLs from the LMC, SMC and Galactic bulge,][]{watson2006,soszynski2014,soszynski2016}. This is in clear contrast with stars of other types. 

The vast majority of these candidates were detected indirectly by analysing cyclic variations of the pulsation period caused by hypothetical orbital motion around center of mass together with an unseen companion -- the well known Light Travel-Time Effect \citep[LTTE,][]{irwin1952,sterken2005a}. In the case of RRLs this method is the most promising because of their physical properties and evolutionary history. If the companion is of lower luminosity, mass or size, it would be hardly detectable via eclipses, radial-velocity (RV) measurements, or colour peculiarity \citep[see the discussion in][]{skarka2016a}. It could be naturally expected that only wide systems with very long orbital periods (typically more than hundred of days) can be detected via LTTE because the changes in time delays are proportional to the size of the orbit. Tight systems with small orbital semi-major axis will cause only very small variations of the pulsation period of RRL that would be hardly detectable.

However, the wide systems identifiable using LTTE are the most important and interesting because the large separation secures that the RRL component is not affected by the mass transfer, which allows us to investigate a classical RRL that does not differ from an RRL star that evolved separately as an isolated single star. Evolution of stars in short-period semi-detached or contact systems can also produce stars that behave like a classical RRLs \citep[the so called Binary Evolution Pulsators -- BEPs,][]{pietrzynski2012,smolec2013}, but has different physical origin and characteristics. \citet{karczmarek2017} showed that BEPs, that could contaminate about 0.8\,\% of genuine RRL stars, occur only among binaries with orbital periods shorter than about 2000\,d. The vast majority of the recently discovered candidates \citep{hajdu2015,liska2016b} have proposed orbital periods significantly longer than this limit (years to decades) suggesting that classical RRLs could be present in these candidate systems. 

This is also the case of Z CVn (=GSC 03023-00942; J2000 $\alpha =12^{\mathrm{h}}49^{\mathrm{m}}45^{\mathrm{s}}$\hspace{-0.3em}.4, 
$\delta =+43^{\circ}46'25''$\hspace{-0.3em}.1), which is the subject of this paper. Z CVn is an RRab type star discovered by \citet{ceraski1911}. Besides the long-term cyclic period variations identified by \citet{firmanyuk1982} and \citet{leborgne2007}, Z CVn shows significant short-period light curve modulation \citep[the so-called Blazhko effect,][]{blazhko1907} with a period of about $22.75$\,d \citep{kanyo1966}. The phenomenon in this star is even more interesting since the modulation period was reported to change with opposite sign to changes in the pulsation period \citep{leborgne2012}. 

We performed extensive photometric (sect. \ref{Sect:Photometry}) and spectroscopic monitoring (sect. \ref{Sect:Spectroscopy}) of Z CVn during four and two seasons, respectively. We utilized available maximum times from the literature and also collected available data from various sky surveys to get as many maximum times as possible for the investigation of cyclic long-term period changes during the past century (sect. \ref{Sect:PeriodEvolution}). In this section we also model the possible binary orbit. In sect. \ref{Sect:Blazhko effect} we investigate the period variations on shorter time scales and investigate the Blazhko behaviour. The basic physical characteristics and their variations during the Blazhko cycle are investigated in sect. \ref{Sect:PhysChar}. Thorough discussion about the binarity is given in sect. \ref{Sect:BinaryHypothesis}. We summarize the results in sect. \ref{Sect:Discussion}.

\section{Photometric observations and maximum times}\label{Sect:Photometry}

\begin{figure*}
	\includegraphics[width=2\columnwidth]{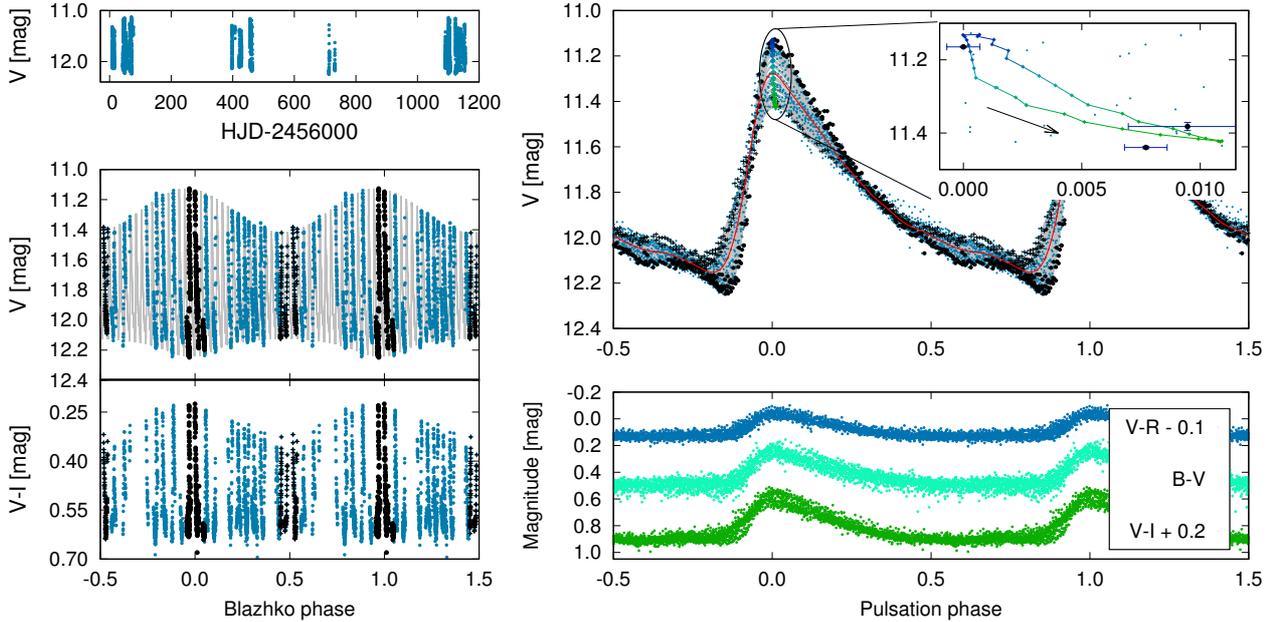}
    \caption{Distribution of the photometric data (top left), data phased with Blazhko and pulsation period (bottom left and top right panel, respectively). The bottom right panel shows the colour indices phased with the pulsation period. Blazhko phases around 0.0 and 0.5 are highlighted with black symbols. The grey dots show the synthetic light curve according to our model (see sect. \ref{Sect:Blazhko effect}). The magnified part around the maximum light in the top right panel shows synthetic maxima of the light curve during one modulation cycle (the direction of the movement is shown by the arrow). Three maxima from 2012 are shown with black circles. The full red line shows the mean light curve.}
    \label{Fig:Phases}
\end{figure*} 
   
Z CVn was observed in the frame of \textit{the Czech RR Lyrae stars observing project} \citep{skarka2013a,skarka2015a} during 2012--2015 at a private observatory in Znojmo, Czech Republic (see the data distribution in the top panel of Fig. \ref{Fig:Phases}). Multicolour $BVR_{\rm{c}}I_{\rm{c}}$ 120s exposures were gathered using an 8-inch Schmidt-Cassegrain telescope equipped with G2-0402 CCD camera (768$\times$512 KAF-0402ME chip). Observations from Znojmo were complemented in 2015 with data gathered at the Observatory and Planetarium Brno (14-inch Schmidt-Cassegrain telescope equipped with G4-16000 CCD camera, KAF-16803 chip, 5 nights). In total, about 2400 points in each filter were collected during 53 nights spanning almost 1150 days\footnote{All the data are available on-line as a supporting material to this paper.} (for more details see Table \ref{Tab:Obslog} with the journal of observations).

\newcolumntype{M}{>{\footnotesize}c}
\begin{table}
\centering
\caption{Observation log. The last four rows show number of frames in particular filters.}
{\footnotesize
\def\arraystretch{1.1}
\tabcolsep=3pt
\begin{tabular}{llllllll}
\hline\hline
Season  & 2012 & 2013 & 2014 & 2015 & Total \\
Nights  & 21 & 12 & 3 & 17 & 53 \\
Start   & 2456008 & 2456397 & 2456712 & 2457090 & - \\
Span [d]& 68 & 66 & 19 & 64 & 1146 \\
 $B$ 	& 968 & 428 & 82 & 961 & 2439 \\
 $V$ 	& 954 & 468 & 83 & 951 & 2456 \\
 $R_{\rm{c}}$ & 957 & 443 & 82 & 935 & 2417 \\
 $I_{\rm{c}}$ & 936 & 428 & 81 & 942 & 2387 \\

\hline
\end{tabular}\label{Tab:Obslog}
}
\end{table}

Each observation was reduced in a classical way using dark and flat field frames. We used \textsc{C-Munipack} software\footnote{\url{http://c-munipack.sourceforge.net/}} for the data reduction and for aperture photometry. USNO-B1.0 1339-0234455 and USNO-B1.0 1338-0234449 served as comparison and check stars (see Table~\ref{Tab:Comparisons}). Relative accuracy of a photometric point was between 0.009 and 0.07\,mag depending on used filter and observing conditions. The data were transformed to standard Johnson-Cousins system applying transformation coefficients determined on the basis of observation of \citet{landolt1992} fields in each season. Standard $BVri$ magnitudes of the comparison stars were adopted from the UCAC4 catalogue \citep{zacharias2013}. Johnson-Cousins {\it R$_{\rm c}$} and {\it I$_{\rm c}$} magnitudes were calculated from {\it r} and {\it i} Sloan magnitudes using equations from \citet{jordi2006}.

Based on our photometric data, we estimated the light elements as
\begin{equation}\label{Eq:PulsElements}
   	{\rm HJD}~T_{\rm{max}} = 2456048.3301(7) + 0\overset{\rm d}.6539498(3)\times E_{\rm{p}}.
\end{equation}
These values were used for construction of the pulsation-phase curves shown in Fig. \ref{Fig:Phases}. The zero epoch from the eq.~\ref{Eq:PulsElements}, which was chosen as the best defined among the maxima with the highest amplitude, was also used as the basic zero phase for blazhko-phased curves. For more details about the period investigation, methods and software used see sect. \ref{Sect:PeriodEvolution} and \ref{Sect:Blazhko effect}.

\newcolumntype{M}{>{\footnotesize}c}
\begin{table}
\centering
\caption{Stars used as comparison and check stars (USNO-B1.0~ID).}
{\footnotesize
\def\arraystretch{1.1}
\tabcolsep=3pt
\begin{tabular}{lll}
\hline\hline
ID & 1339-0234455 & 1338-0234449 \\
RA (2000) & 12 49 50.3 & 12 50 21.9 \\
DEC (2000)& +43 54 19.6 & +43 54 19.6 \\
$B$\,[mag]& 12.770 & 13.457 \\
$V$\,[mag]& 12.037 & 12.743 \\
$R_{\rm{c}}$ [mag] & 11.721 & 12.387 \\
$I_{\rm{c}}$ [mag] & 11.326 & 11.897 \\
\hline
\end{tabular}\label{Tab:Comparisons}
}
\end{table}

From the well-covered maximum phases present in our $V$ dataset we determined 18 maximum times using polynomial fitting method described in \citet{skarka2015b}. Moreover, we got three additional maximum times (in 2013 in Str\"{o}mgren $y$ and in 2016 in $V$) with different telescopes with unknown calibration coefficients\footnote{This is the reason why these observations are not included in our photometric data set.}. New maximum times were complemented with maximum times from the GEOS RR Lyr database\footnote{\url{http://rr-lyr.irap.omp.eu/dbrr/index.php}} \citep[version July 19, 2016,][]{boninsegna2002,leborgne2007}. We also searched the literature for maximum times not listed in the GEOS database, and determined new ones from various sky surveys. This gave us 19 values from \citet{prager1939}, 1 value from \citet{blazhko1926}, 1 value from \citet{strelkova1964}, and 7 values from \citet{lampe1970}, 6 determined from SuperWASP survey \citep{pollacco2006,butters2010}, 6 from the NSVS (ROTSE) data \citep{wozniak2004}, 11 from the DASCH project \citep[e.g.][]{grindlay2009}.

Maximum times from sources with sparse data (NSVS, DASCH) were calculated as the mean maximum time from 30 points using template fitting method described in \citet[][]{liska2016a}. We omitted data with high uncertainty and took into account various length of exposures of the photographic plates in case of the DASCH data \citep[for details see][]{liska2016a}. From the GEOS database we omitted maximum times marked as \textquoteleft pr. com.\textquoteright, a few other values with unfindable source, and apparent outliers. All in all we used 227 maximum times (155 from GEOS database), out from which 57 were photographic, 53 visual+unknown origin (51+2)\footnote{The two maximum times with unknown origin are probably based on visual observations as noted in the GEOS database, but we did not succeed in finding the original papers by N.~Florja cited in \citet{prager1939} where the maximum times come from.}, 14 photoelectric, 103 CCD+DSLR (101+2). Although our sample covers more than 120 years\footnote{ The first maximum time comes from Harvard plate collections and dates back to 1894 \citep{prager1939}.} and contains a lot of maximum times, it cannot be considered as complete. For example, we did not succeed in finding original maximum times that \citet{firmanyuk1982} used to plot \oc~in his fig. 4. All maximum times that we used are available on-line as a supporting material to this  paper and at the CDS.

\section{Spectroscopy and radial velocity}\label{Sect:Spectroscopy}

Z CVn was spectroscopically observed in 2015 and 2016 as one of the primary targets of the observing project focused on binarity among RR~Lyrae stars announced by \citet{guggenberger2016}. Observations were carried out at McDonald Observatory, Texas using $f$/13.5 Cassegrain focus of the 2.1-m Otto Struve telescope. Spectra were gathered with the Sandiford Echelle Spectrometer \citep{mccarthy1993}. We have continuous spectral coverage over the observed range of 4250--4750\,\r{A} with resolving power of $R\sim55\,000$. Integration time was 1690 seconds for all observations (3\,\% of the pulsation period). Th- Ar spectra were taken for wavelength calibration after each stellar spectrum, and before the telescope was moved. The signal to noise ratios per resolution element are between 10 and 33, depending on the brightness of the star and the sky conditions.

We used standard IRAF\footnote{IRAF is distributed by the National Optical Astronomy Observatories, which are operated by the Association of Universities for Research in Astronomy, Inc., under cooperative agreement with the National Science Foundation.} routines for data reduction. Radial velocity standard stars were observed several times during the course of each night from the Geneva list of CORAVEL standard stars\footnote{\url{http://obswww.unige.ch/~udry/std/stdcor.dat}}.  The radial velocities were  determined by cross-correlating the standard star spectra with the Z CVn spectra using the IRAF task fxcor. Only metal lines were used in the cross-correlation, as the hydrogen lines are known to have a phase offset and different velocity amplitudes compared to metal lines \citep{sanford1949,oke1962,sesar2012}. Eleven RVs estimated from our observations are given in Table \ref{Tab:RVtable}.

\begin{table}
\centering
\caption{Newly determined and publicly available RV measurements for Z~CVn.}
{\footnotesize
\def\arraystretch{1.1}
\tabcolsep=3pt
\begin{tabular}{llllll}
\hline\hline
\multicolumn{1}{c}{$T_{\rm obs}$ [HJD}& RV&Ref.&\multicolumn{1}{c}{$T_{\rm obs}$ [HJD}& RV &Ref.\\
\multicolumn{1}{c}{$-$2400000]}&[km/s]&&\multicolumn{1}{c}{$-$2400000]}&[km/s]&\\ \hline
57087.7387& +6.3(1.9)&TS&57409.007& +15.0(4.7)&TS \\
57087.8201& +14.6(1.6)&TS&22884.743&+8.3(35)&A73\\
57087.9056& +18.2(1.4)&TS&24185.024&+34(35)&A73\\
57091.8589& +18.2(1.9)&TS&45460.721&+29(12)&HB85\\
57091.94& +10.2(0.9)&TS&45460.7345&+15(12)&HB85\\
57092.7161& $-$25.3(1.0)&TS&45462.6765&+7(12)&HB85\\
57176.6637& +1.4(1.1)&TS&45462.69&+14(12)&HB85\\
57177.6545& $-$7.1(2.4)&TS&47973.8224& $-$28(28) &L93, L94\\
57177.7476& $-$14.3(2.0)&TS&47974.879& +34(32) &L93, L94\\
57408.9802& +13.9(1.3)&TS&&&\\
\multicolumn{6}{l}{{\bf Notes:} A73 -- \citep{abt1973}; HB85 -- \citep{hawley1985};}\\ \multicolumn{6}{l}{L93 -- \citep{layden1993}; L94 -- \citep{layden1994}.; TS -- This Study}\\
\hline
\end{tabular}\label{Tab:RVtable}
}
\end{table}

Besides our measurements we searched for the RVs in literature. The first RV measurement comes from \citet{joy1938}. Unfortunately, his value was published without observing date. Other RV comes from  \citet[][no time of observation given]{preston1959} who detected changes in spectral type during pulsation cycle as A8--F5 (H$\gamma$-line) and A3--A7 (Ca II K-line). Furthermore we found only three sources of RV measurements -- 2~RVs from \citet{abt1973}, 4~RVs from \citet{hawley1985}, and 2~RVs from \citet{layden1993,layden1994}\footnote{He published only the mean RV. Therefore, we used measurements from his PhD thesis available at \url{http://physics.bgsu.edu/~layden/ASTRO/PUBL/THESIS/}.}. All used RV values are listed in Table \ref{Tab:RVtable}.

Radial-velocity datasets from \citet[][observed in 1921 and 1925]{abt1973} were published without individual uncertainties, but \citet{layden1994} assign the measurements taken in the first half of the 20th century with common uncertainty of 35\,km/s. Measurements by \citet[][also without individual uncertainties]{hawley1985} have common uncertainties of 12\,km/s, accuracy of \citet{layden1993,layden1994} is within 30\,km/s. Our measurements are of an order of magnitude better precision than these measurements (typically about 2\,km/s).

The problem of using historical measurements is that the RVs were estimated on the basis of different spectral lines that form in different layers of the stellar atmosphere. Thus, the combination of historical measurements could be tricky. \citet{abt1973} does not provide information about used lines, \citet{layden1993,layden1994} used H$\beta$ and H$\gamma$ Balmer lines and Ca II K-line, while \citet{hawley1985} and we used only metallic lines. To unify the heterogeneous RVs estimations and determine the correct systemic RV we used the RV templates from \citet{sesar2012}, who provides normalised RV curves for H$\alpha$, H$\beta$, H$\gamma$ and metals.

We followed the approach described in \citet{liska2016a}. First we modelled the template from metallic lines adopted from \citet{sesar2012} with high-order harmonic polynomial and compared them with the four observational data sets (Table \ref{Tab:RVtable}) that were time-shifted according to cyclic period changes (subtraction of the model of the LTTE, sect. \ref{Sect:LTTE}). Individual systemic RV shifts ($\gamma$ velocity) for each data set (Table \ref{Tab:RVgvelTable}) were estimated using non-linear least-squares method. Uncertainties were determined statistically using bootstrap-resampling method.  

\begin{table}
\centering
\caption{Newly determined systemic RVs of Z~CVn (third column), their errors (the last column) and the mean date of the measurement (the second column). The short cuts of the sources are the same as in Table~\ref{Tab:RVtable}.}
{\footnotesize
\def\arraystretch{1.1}
\tabcolsep=3pt
\begin{tabular}{lccc}
\hline\hline
Source & $T_{\rm mean}$ & $\gamma$ [km/s] & Err $\gamma$ [km/s] \\ \hline
A73 & 2423535 & 8.1 & 13.0 \\
HB85 & 2445462 &	10.6 &	4.1 \\
L94 & 2447975 &	1.1	& 12.1 \\
TS & 2457172 &	$-0.9$	& 2.2 \\
\hline
\end{tabular}\label{Tab:RVgvelTable}
}
\end{table}

\section{Long-term period evolution}\label{Sect:PeriodEvolution}

Already \citet{blazhko1922} noticed from his 25 maximum times determined between 1912 and 1917 that the pulsation period of Z CVn varies. Four years later, \citet{blazhko1926} published the light elements that included quadratic term indicating slow lengthening of the pulsation period. \citet{firmanyuk1980a,firmanyuk1980b,firmanyuk1982} was the first who showed that Z~CVn undergoes strong cyclic period variations with period of about 60\,years and amplitude of the \oc~of about 0.6\,d. Similar cyclic pattern as the one shown in \citet{firmanyuk1982} also shows the \oc~diagram in \citet[][fig. 4]{leborgne2007}. 

\begin{figure*}
	\includegraphics[width=1.33333\columnwidth]{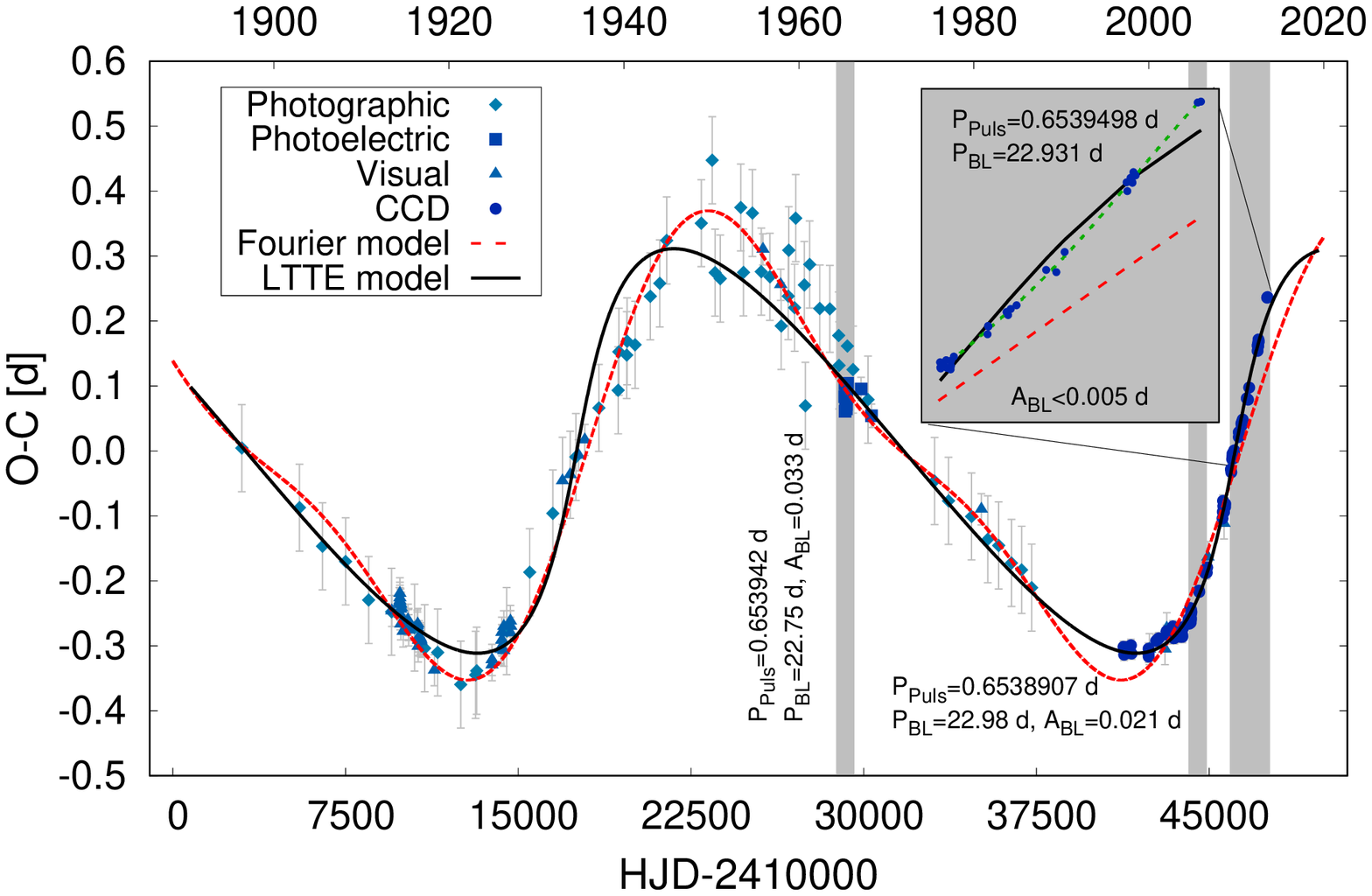}\includegraphics[width=0.66667\columnwidth]{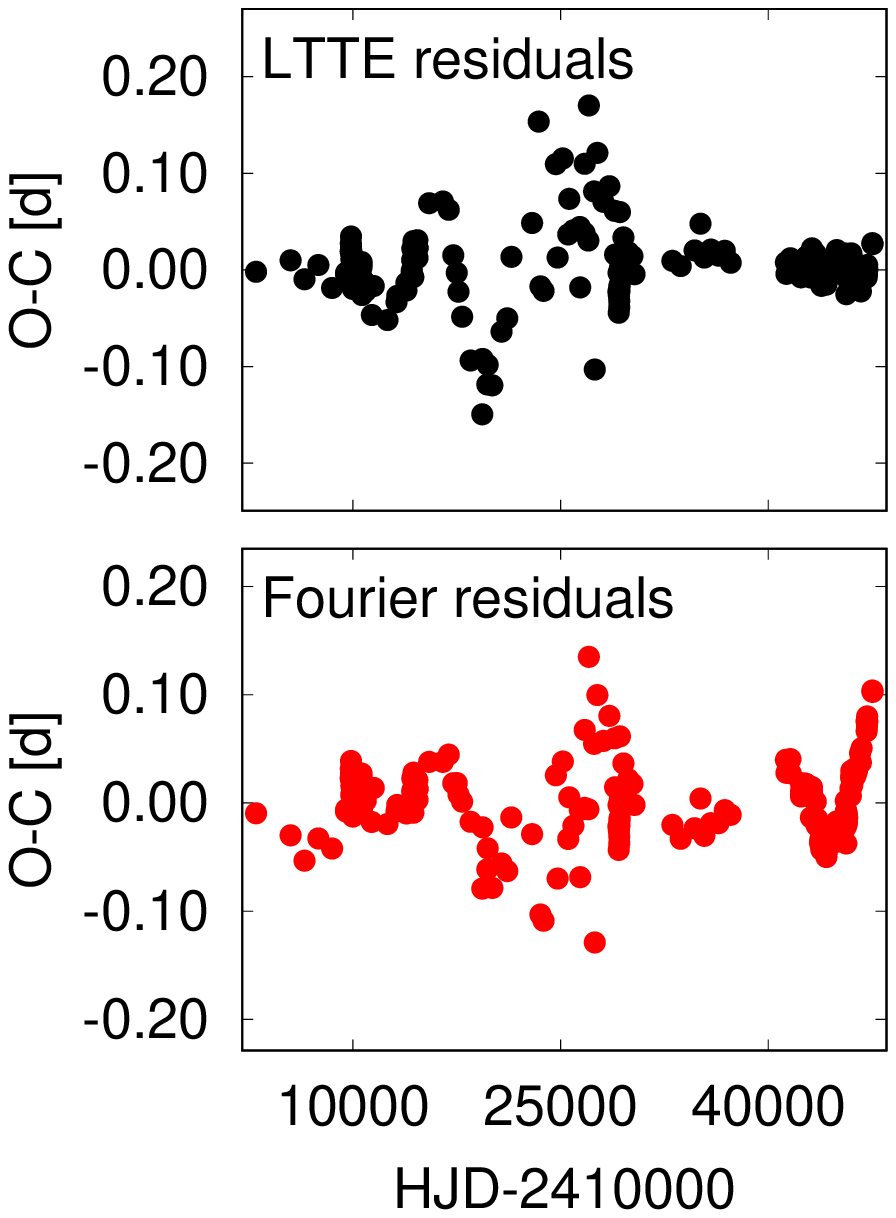}
    \caption{{\it Left:} The overall period evolution over the time span of more than 100 years (the main panel). Different symbols denote different types of observations (diamonds - photographic, squares - photoelectric, triangles - visual, circles - CCD). The full black line shows the model based on the LTTE assumption, while the red dashed line shows the Fourier fit with two sine components. The gray-shaded areas show the intervals when pulsation and modulation periods and modulation amplitude were determined (see the text for more details). {\it Right:} Residuals after removing the models.}
    \label{Fig:LTTE}
\end{figure*}

The \oc~diagram constructed from our set of maximum times also suggests that the pulsation period could be cyclically unstable with period of about 80\,years (the main panel of Fig. \ref{Fig:LTTE}). The amplitude of the period oscillations is huge and almost exceeds the length of the pulsation cycle. Such strong variations are not common among RR Lyrae type stars, but also not unique \citep[see e.g.][]{firmanyuk1982}. However, our understanding of period changes in RRLs is still poor and such strong and possibly cyclic behaviour is difficult to explain.

The stellar evolution and pulsation theory suggest that RRLs should slowly change their pulsation periods during evolution through the instability strip simply because of slow change of their radius according to the period-mean density law $P\sqrt{\rho}\sim$\,const \citep{ritter1879}. This smooth secular period changes, which can go in both directions (period shortens when the star evolves blueward, while it lengthens when the star evolves towards the red edge of the instability strip), are commonly observed in globular clusters \citep[e.g. $\omega$ Cen, IC4499, M5, M3,][just to mention some recent studies]{jurcsik2001,kunder2011,szeidl2011,jurcsik2012a}, but also in field stars \citep[e.g.][]{leborgne2007}. However, significant portion of stars investigated in these studies show erratic abrupt changes \citep[in the Galactic field it is e.g. XZ Cyg or RR Gem,][]{lacluyze2004,sodor2007} and/or rates of period changes that are an order of magnitude larger than evolutionary effects could explain \citep{iben1970}. This means that there must be some additional effects that act in RRL stars that could cause complex period evolution on a significantly shorter than evolutionary time scales. It was found that \bl~stars show complex/irregular \oc s with large amplitudes more often than stars with stable light curve, thus, suggesting that \bl~effect and rapid irregular period changes could have a common origin \citep{jurcsik2012a,szeidl2011}.

\citet{sweigart1979} suggested that the discrete mixing events near the boundary of semiconvective zone of an RRL star could be responsible for the positive, negative, but also irregular period changes. Also \citet{firmanyuk1982} supposed that cyclic period variations could rise from changes of the internal structure of a star, but without any further details. However, \citet{arellano2016b} found that, at least for some stars in M5, the irregular \oc~variations could be caused by improper counting of the cycles and that it must not be a real feature of a star. 

An alternative explanation of the short-time scale sudden variations of the pulsation period comes from \citet{cox1998} who suggests that the period changes could be caused by small changes in a helium composition gradient below the hydrogen and helium convection zones. He proposes that the time scales of the variations could correspond to the time scale of an occasional dredge up of helium caused by convective overshooting ($\sim$days).

Another explanation, at this time relying to the origin of the cyclic period variations, counts with the hydromagnetic effects that take place periodically -- analogue to the solar magnetic cycle \citep{stothers1980}. Change in the magnetic energy content during the cycle should result in a change of stars radius that will further affect the pulsation period. This effect was offered as a possible explanation of the cyclic \oc~variations of BE Dor by \citet{derekas2004}. However, as they already noted, this explanation faces significant troubles since it seems that RRLs do not have strong surface magnetic fields \citep{chadid2004}. 


\subsection{The LTTE}\label{Sect:LTTE}

Because none of the proposed hypotheses was supported by observations, the most promising explanation of smooth, regular and (possibly repetitive \oc~variations of Z CVn is the binary scenario that could be modelled with LTTE. Among RRLs the LTTE has been very rarely reported so far and is proposed only in a few dozens of candidates \citep{hajdu2015,liska2016b,liska2016c}.

We used the methods and codes thoroughly described in \citet{liska2016a} to model the \oc~variations and get parameters of suggested binary based on LTTE. During the fitting process the best model is found using non-linear least squares method and uncertainties of the final parameters are estimated statistically via bootstrap resampling (5\,000 repetitions of the calculation with re-sampled datasets). The final model is shown with the full black line in the left panel of Fig. \ref{Fig:LTTE} and pulsation and orbital parameters resulting from the fit are listed in Table \ref{Tab:LiTEparam}. Note that the mean pulsation period during the assumed 80-year cycle is significantly different (of about $10^{-4}$\,d) from the actual period in eq. \ref{Eq:PulsElements}. This demonstrates well how dramatic the variation is.

Because the uncertainties of the maximum times are often missing, unreliable or unrealistic, we assign photographic:visual+unknown:photoelectric:CCD+DSLR (shown with different symbols in the main panel of Fig. \ref{Fig:LTTE}) with weights 1:7.7:14.2:50.4 that are based on the uncertainties of particular type of maxima that were estimated on the basis of scatter in the residual \oc~(0.067:0.024:0.018:0.009\,d). These errors are probably slightly larger than the real uncertainties because of additional scatter that could be present due to the Blazhko effect, which causes (small) variations in the \oc~amplitude (Sect. \ref{Sect:Blazhko effect}).

The large amplitude of the \oc~dependence, which provides information about the size of the orbit ($A_{\rm LTTE}=0.3113$\,d; $a_{1}\sin i/c=69.7$\,au), in combination with a relatively short LTTE~period ($P_{\rm orbit}=28\,590$\,d) suggest that the system must be very special with an exotic companion. If the binary hypothesis is correct, our model (Table \ref{Tab:LiTEparam}) implies that Z~CVn orbits around a massive black hole (BH) with a minimal mass of $\mathfrak{M}_{\rm 2, min} \!\sim\! 57$\,$\mathfrak{M}_{\odot}$ at a very eccentric orbit with $e\!\sim\!0.63$.

Alternatively, the \oc~dependence could be comparatively well described with Fourier series (one frequency and its harmonics) with $P=78$\,years. The dashed red line in the left panel of Fig. \ref{Fig:LTTE} shows the fit with two harmonics.)\footnote{The error of this period is comparable with the value itself.} Residuals after removing the main trend shown in the right-hand panels of Fig. \ref{Fig:LTTE} are still quite large and show similar dependence in both cases suggesting some additional quasi-periodic long cycle. Naturally, the more harmonics are used, the lower the residuals. It is very difficult to decide whether the shape of residuals has real origin, or is simply produced by the inappropriate models. In addition, residuals of the LTTE model after HJD 2453000 (populated mainly with high-quality CCD measurements) differ from the residuals before this date. This is also very suspicious and gives rise to doubts about the relevance of the \oc~models. The binary hypothesis is thoroughly discussed in Sect. \ref{Sect:BinaryHypothesis}.


\begin{table*}
\centering
\begin{minipage}{175mm}
\caption{Pulsation and orbital parameters for the binary hypothesis$^{*}$.}
{\footnotesize
\def\arraystretch{1.5}
\tabcolsep=1.4pt
\begin{tabular}{ccccccccccccc}
\hline\hline
$P_{\rm puls}$			& $M_{0}$			& $P_{\rm orbit}$	& $T_{0}$			& $e$			& $\omega$		& $A$				& $a_{1}\sin i$		& $f(\mathfrak{M})$		& $\mathfrak{M}_{\rm 2, min}$ 	& $K_{1}$ 		& $\chi_{\rm R}^2$	& $N_{\rm max}$\\
$[d]$				& [HJD]					& [d]			& [HJD]				&			& $[^{\circ}]$		& [light day]			& [au]			& [$\mathfrak{M}_{\odot}$]	& [$\mathfrak{M}_{\odot}$]	& [km\,s$^{-1}$]	&			& \\
\hline
0.65384853$^{+11}_{-10}$	& 2453531.6768$^{+44}_{-52}$	& 28590$^{+130}_{-110}$	& 2456149$^{+74}_{-88}$	& 0.6344$^{+98}_{-85}$	& 0.7$^{+1.6}_{-1.8}$	& 0.4027$^{+55}_{-61}$	& 69.7$^{+1.0}_{-1.1}$	& 55.3$^{+2.3}_{-2.6}$		& 56.5$^{+2.3}_{-2.6}$		& 34.33$^{+52}_{-54}$	& 1.074(95)		& 227\\
\hline
\end{tabular}\label{Tab:LiTEparam}
}

{\scriptsize {\bf Notes.} $^{(*)}$ Columns contain following parameters: $P_{\rm puls}$ -- main pulsation period, $M_{0}$ -- zero epoch of pulsations, $P_{\rm orbit}$ -- orbital period, $T_{0}$ -- time of periastron passage, $e$ -- numerical eccentricity, $\omega$ -- argument of periastron, $A$ -- $a_{1}\sin i$ in light days (semi-amplitude of \lt~$A_{\rm LTTE}$ can be calculated as $A_{\rm LTTE} = A\,\sqrt{1-e^{2}\,\cos^{2}\omega}$), $a_{1}\sin i$ -- projection of semi-major axis of primary component (RRL) $a_{1}$ according to the inclination of the orbit $i$, $f(\mathfrak{M})$ -- mass function, $\mathfrak{M}_{\rm 2, min}$ -- the lowest mass of the second component, the value was calculated for inclination angle $i=90^{\circ}$ and adopted mass of primary $\mathfrak{M}_{1} = 0.6$\,$\mathfrak{M}_{\odot}$, $K_{1}$ -- semi-amplitude of RV changes primary component, $\chi_{\rm R}^2$ -- normalised value of $\chi^{2}$, where $\chi^{2}_{\rm R} = \chi^{2}/(N_{\rm max}-g)$ for number of used maxima times $N_{\rm max}$ and number of free (fitted) parameters $g$ (LTTE, $g=7$), $N_{\rm max}$ -- number of used maximum times.} 
\end{minipage}
\end{table*}

\section{Short-term period evolution and the Blazhko effect}\label{Sect:Blazhko effect}

\begin{figure}
	\includegraphics[width=\columnwidth]{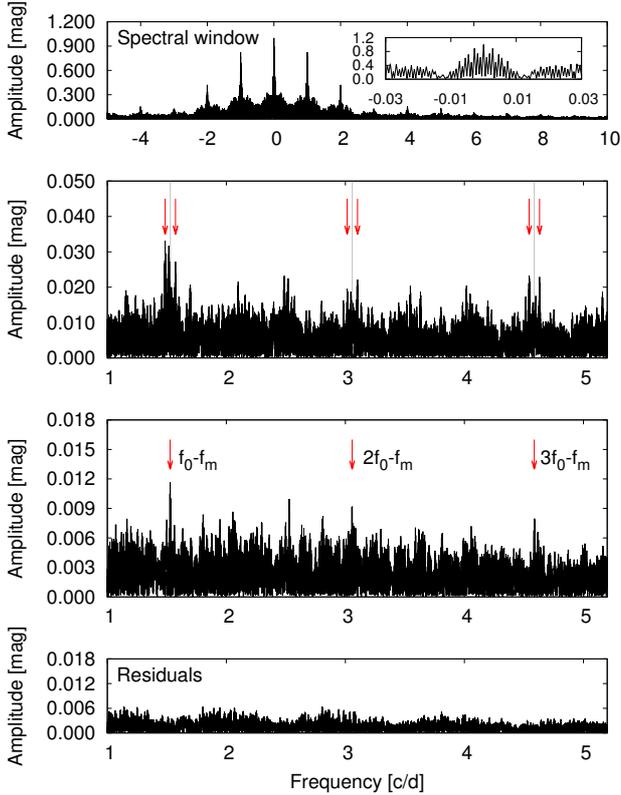}
    \caption{Frequency window (the top panel) together with the frequency spectra during the prewhitening process. The second panel from the top shows the frequency spectra after prewhitening with the basic pulsation components. The positions of the side peaks are marked with arrows. The bottom middle panel shows the situation after the removal of the side peaks (arrows show the position of the side peaks related to additional long-term modulation). The bottom panel shows the residuals after prewhitening with all frequencies listed in Tab. \ref{Tab:Freq}.}
    \label{Fig:FrekSpec}
\end{figure}

The amplitude and phase/frequency modulation of the light curve of Z CVn is known since the study by \citet{blazhko1922}. From all panels of Fig. \ref{Fig:Phases} it is apparent, that the modulation is quite strong and dominates in amplitude which exceeds almost 0.3\,mag in $V$. Because our data are single-site and show strong seasonal pattern (the top left panel of Fig. \ref{Fig:Phases}), the frequency spectra are strongly dominated by daily and yearly aliases (see the spectral window shown in the top panel of Fig. \ref{Fig:FrekSpec}).

\textsc{Period04} \citep{lenz2005} was used for consecutive prewhitening and proper identification of the significant peaks with $S/N>3.5$\footnote{The $S/N$ is calculated as the ration between the amplitude of the peak and average amplitude of the frequency spectra in the $\pm1$\,c/d interval around the peak.}. We analysed the data in all passbands separately with permanent visual supervision at all steps, which was necessary to prevent confusion with aliases. As expected, we identified the basic pulsation frequency $f_{0}$ and its harmonics to the 10th order in {\it V} and {\it R$_{\rm c}$} (see Table \ref{Tab:Freq} with the full frequency solution in all four filters). In addition, close equidistant peaks near the pulsation harmonics $kf_{0}\pm f_{\rm BL}$, $kf_{0}- f_{\rm M}$ were identified (red arrows in the middle panels of Fig. \ref{Fig:FrekSpec}). Such peaks are assumed to be the products of modulation with frequency that is equal to the spacing between pulsation harmonic and side-peak \citep{szeidl2009,szeidl2011}. No peak corresponding directly to the modulation frequency, which is usually present in the frequency spectra of high-quality data of \bl~stars, was detected. Because this peak relates to the change of the mean magnitude \citep{szeidl2011}, the mean magnitude varies by less than 0.006\,mag, which is the significance limit in the region where the assumed peak should be situated. 

When the list with significant peaks was available, we used the LCfit routine \citep{sodor2012a} and modelled the light changes using
\begin{equation}\label{Eq:LCModel}
A(t)=A_{0}+\sum^{N}_{i=1}A_{i}\sin \left( 2\pi f_{i}[t-T_{0}]+\varphi_{i}\right),
\end{equation}
where $A_{i}$ and $\varphi_{i}$ is the amplitude and phase of the $i$-th component with frequency $f_{i}$, $T_{0}$ is the zero epoch from eq.~\ref{Eq:PulsElements}, and $N$ is the number of the relevant detected peaks. Because the frequencies are not independent, we adjusted only \f0, \f0$-$\fbl~and \f0$-$\fm, the other frequencies were calculated from these three independent frequencies. 

The values of frequencies and corresponding pulsation and modulation frequencies slightly differ for each passband (within $2\sigma$). Because there is no systematics from short- to long-wavelength filters and different period in different filters is also physically impossible, we assume that the differences result from different quality and point distribution in different filters. Thus, we calculated the periods as the average values from the {\it BVR$_{\rm c}$I$_{\rm c}$} data sets. The resulting pulsation, Blazhko and second modulation periods are $P_{\rm puls}=0.6539498(3)$\,d, $P_{\rm BL}=22.931(4)$\,d and $P_{\rm M}=1100(20)$\,d. The long $P_{\rm M}$ could be an artefact of the secular period change, because it is comparable with the time span of the data and corresponding side peaks always appear on the low-frequency side of the pulsation harmonics. 

\begin{figure}
	\includegraphics[width=\columnwidth]{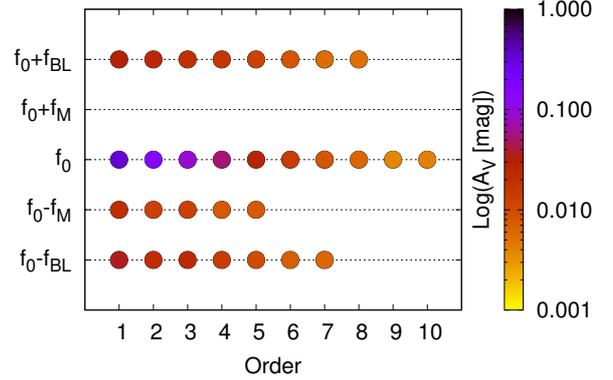}
    \caption{The diagram showing the structure and amplitude of the side peaks around ten detected harmonic orders of \f0. The spacing of the side peaks is not in scale.}
    \label{Fig:Echelle}
\end{figure}

The structure of the side-peak multiplets is apparent from the echelle graph in Fig. \ref{Fig:Echelle}. The highest side peak has amplitude ten times lower than \f0 and the lowest significant peak has amplitude of about 0.004\,mag. It is apparent that the closer peak corresponding to the longer modulation/secular period change was detected only on the left-hand side of \kf0. The amplitudes of the \bl~side peaks are almost equal at all orders\footnote{This further indicates the dominance of the amplitude modulation \citep{benko2011}.}, but at some \kf0 the peak to the right has larger amplitude, while at other \kf0 the peak to the left is higher. Because the amplitude of the peaks can be influenced by the sampling and gaps in the data \citep{jurcsik2005a} we assume that the effect of slightly different amplitude ration of the side peaks is caused by the data characteristics.

\begin{figure}
	\includegraphics[width=\columnwidth]{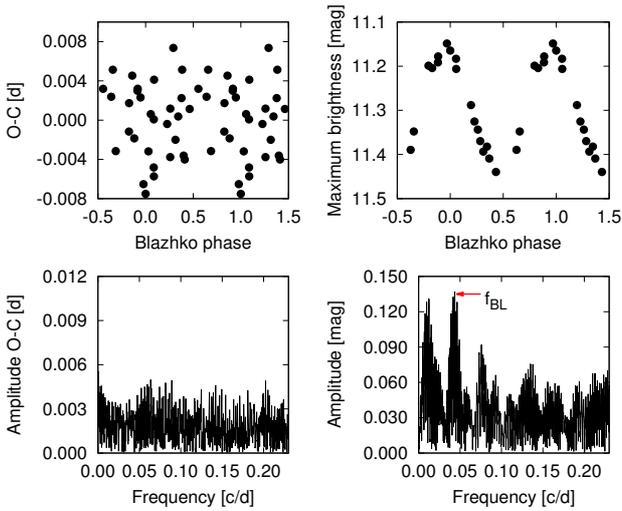}
    \caption{Brightnesses of the maximum light and \oc~values phased with the detected Blazhko period $P_{\rm BL}=22.931$\,d (the top panels) and frequency spectra of the \oc~and maximum magnitudes.}
    \label{Fig:MaxPhased}
\end{figure}

The modulation is well detectable also in the variation of the maximum brightness from our observations ($P_{\rm BL}=22.932(7)$\,d, the bottom right-hand panel of Fig. \ref{Fig:MaxPhased}). This $P_{\rm BL}$ agrees perfectly with the value that we got from the spacing of the side-peaks in the Fourier spectra of the time series. On the other hand, no significant periodicity close to \pbl~with amplitude larger than about 0.005\,d is detectable in CCD \oc s since 2012. It is directly apparent from the amplitude of the residuals after HJD 2450000 in the LTTE residuals in the left panel of Fig. \ref{Fig:LTTE}, and also from the top left panel of Fig. \ref{Fig:MaxPhased} which shows \oc~after removing the long term trend (linear trend + sinusoid with period of about 2300\,d) shown with green dashed line in the insert in Fig.~\ref{Fig:LTTE}. The full amplitude of the phase changes during the \bl~cycle corresponds to about 0.011\,d (see the insert in Fig. \ref{Fig:Phases}), which is the full amplitude in the \oc\ shown in the top right-hand panel of Fig. \ref{Fig:MaxPhased}. No wonder that no periodicity is detected in the corresponding frequency spectra (bottom left panel of Fig.~\ref{Fig:MaxPhased}).

However, based on light-curve fitting in different \bl\ phases we detected phase variations of the basic pulsation frequency peak that correspond to \oc\ variations with amplitude 0.01\,d (see Sect. \ref{Subsect:IPM} and the second top plot in the left panel of Fig. \ref{Fig:IPM}). Possible explanation is that maximum time takes into account only data points around light maximum, thus it has larger uncertainty than a Fourier phase that relies on the data from the full pulsation cycle. 

A very weak period modulation during \bl~cycle is in contrast to \citet{leborgne2012} who clearly detected  also phase modulation with the peak-to-peak amplitude of 0.021\,d and period $P_{\rm BL}=22.98$\,d in their TAROT observations (see their fig. 7). They found that in \citet{kanyo1966} data around JD 2439000 the pulsation period was 0.653942\,d, modulation period was 22.75\,d and full amplitude of the \oc~variations was 0.033\,d. In TAROT data (JD 2454200) the pulsation period was shorter (0.6538907\,d), Blazhko period longer (22.98\,d) and the full \oc~amplitude was 0.021\,d (see the shaded areas and corresponding labels in the main panel of Fig. \ref{Fig:LTTE} to get a better picture).

At the time of our observations (after JD 2456000) the pulsation period roughly exceeds the period in the Kany\'{o}'s time, but the Blazhko period is longer than was around JD 2439000 and the amplitude of the \oc~is so low that the modulation in period is undetectable. It is likely that the variation of the Blazhko period somehow follows the variation of the pulsation period, but since we have only three estimates (at Kany\'{o}s, Leborgne's and our times) we cannot say much about their correlation. If the \oc\ is periodic, it would be interesting to investigate the relation of the pulsation and modulation periods after the pulsation period starts to lengthen (after about 2462000). It is also interesting to point out that the \pbl/\ppuls~ratio oscillates around 35 and the actual value is very close to this value (\pbl/\ppuls$\sim35.065$).

\section{Physical Characteristics}\label{Sect:PhysChar}

Unfortunately, our low {\it S/N} spectroscopy does not allow us to estimate the atmospheric parameters directly from comparison with theoretical spectrum. In addition, the spectra change during the \bl~cycle and we have only very sparse observations. However, there is another simple way to get information about the basic physical parameters from photometry.

\begin{figure*}
	\includegraphics[width=2\columnwidth]{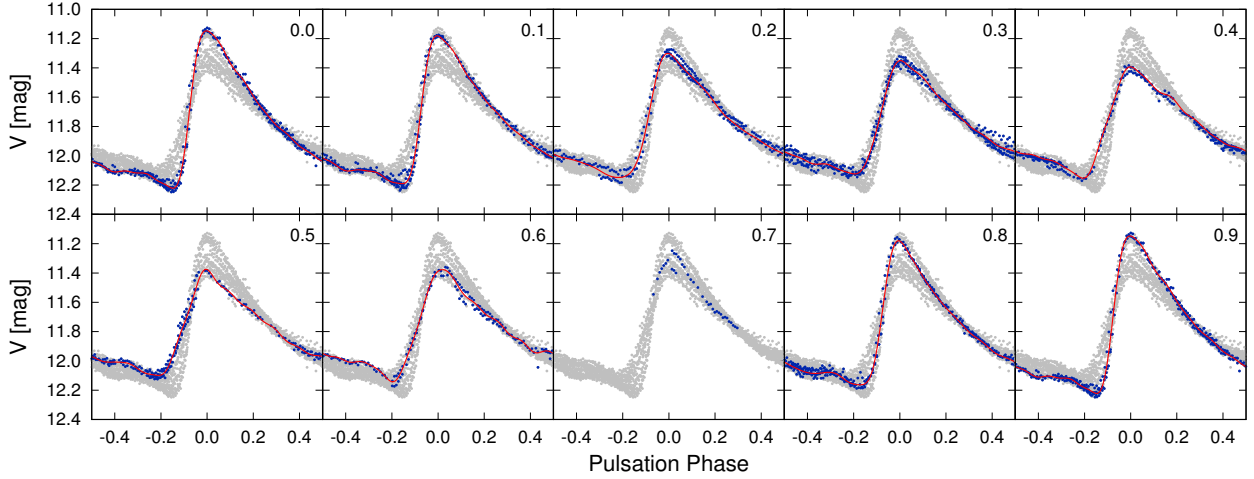}
    \caption{Data for particular \bl~phases (blue points) folded with pulsation period. Grey dots show phase curves of all our $V$-observations. The model light curve is shown with the full red line for each \bl~phase.}
    \label{Fig:BLPhased}
\end{figure*}

Because the light curve characteristics necessarily reflect the physical conditions inside the star, it is possible to estimate them just by investigation of the shape of the light curve. This approach employing phase-independent Fourier parameters \citep{simon1981} was applied many times and many calibrations were defined \citep[e.g.][]{jurcsik1996,jurcsik1998,kovacs2001,nemec2013}. We estimated the parameters in ten different \bl~phases (Fig. \ref{Fig:BLPhased}, Tab. \ref{Tab:PhysChar_App}), as the global value using the mean light curve shown in the top right-hand panel of Fig.~\ref{Fig:Phases}, and using Blazhko-free light curve\footnote{The modulation was filtered out using the light-curve model with parameters in Tab. \ref{Tab:Freq}.} after the modulation was subtracted. The averaged parameters over the \bl~cycle are within the errors the same as the parameters coming from the global mean light curve (compare values for `Average' and `Mean LC' rows in Tab. \ref{Tab:PhysChar_App}), which confirms the already known finding by \citet{jurcsik2009a}. However, we prefer values based on modulation-free light curve because such light curve gives metalicity almost identical to the spectroscopic one (-1.92 in \citet{zinn1984} scale vs. -1.98 estimated by \citet{layden1994}). The details about the procedure and used empirical calibrations can be found in Appendix, sect.~\ref{Sect:PhysChar_App}.


\subsection{Inverse Photometric Baade--Wesselink Analysis}\label{Subsect:IPM}

The inverse photometric Baade--Wesselink method (IPM -- \citet{sodor2009a}) was developed to derive global mean physical parameters of RRab stars from multi-colour photometry, without spectroscopic RV measurements, as well as changes of these parameters during the pulsation and the modulation cycles. The method was successfully applied to several \bl\ stars before \citep{jurcsik2009b,jurcsik2009c,sodor2009b,sodor2011,jurcsik2012b,sodor2012c} to determine the variations in pulsation-phase-averaged mean radius, $\log g$, $M_{V}$, luminosity and effective temperature. During the present study, we employed the IPM on Z~CVn.

We followed the methodology detailed e.g. in \cite{jurcsik2009c}. As a first step, we attempted to determine the Blazhko-phase-independent constant parameters, mass ($\mathfrak{M}$), distance ($d$) and absolute visual magnitude ($M_{V}$). For this reason, we run the IPM on the Blazhko-free light curve of Z~CVn with four different settings (see details in \citealt{sodor2009a}). We accepted metallicity of [Fe/H] = -1.64 (Table \ref{Tab:PhysParam}) for Z~CVn. Unfortunately, when all unknown parameters were free during the fitting process, the method resulted in implausible high $\mathfrak{M}$, $d$ and $M_{V}$ values (about 1.1\,--\,1.4 $\mathfrak{M}_\odot$, 2300\,--\,2600 pc and 0.0\,--\,$-0.25$ mag, respectively). Therefore, we fixed the mass to the value from Table \ref{Tab:PhysParam}. We also fixed the distance at 1760\,pc, $A_{V}=0.05$\,mag and $E(B-V)=0.015$\,mag was taken into account when running the IPM. 

\begin{figure}
	\includegraphics[width=1.1\columnwidth]{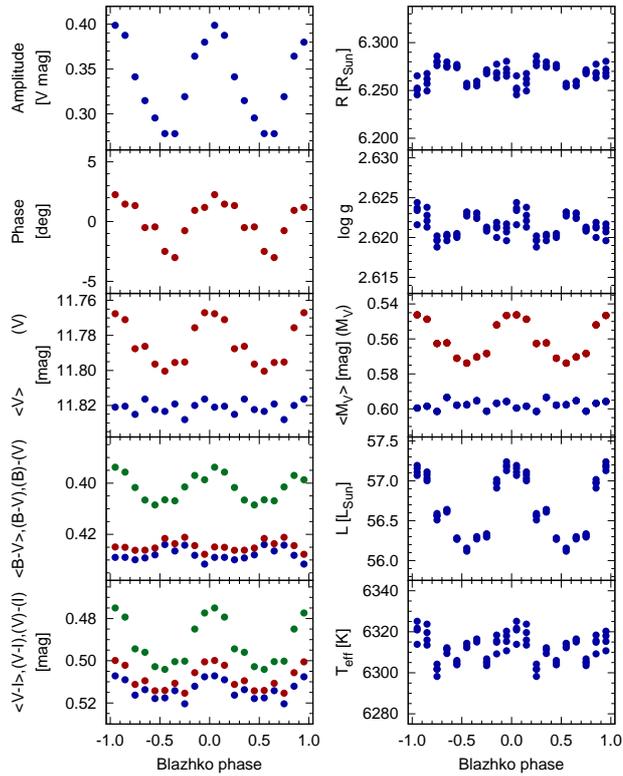}
    \caption{Variation of various parameters during the Blazhko cycle from the IPM method.}
    \label{Fig:IPM}
\end{figure}

When the modulation-free values were determined, we proceeded with the analysis of the different Blazhko phases. To attain sufficient pulsation-light-curve coverage, we divided the data to 10 overlapping Blazhko-phase-bins, extending the bins by $\pm0.075$ Blazhko phase towards both directions, that is, each bin had a width of 0.25 in modulation phase. Even though 4 equal-length non-overlapping bins would be completely independent and sufficient to resolve the variations with the Blazhko modulation, we chose to show the results for 10 overlapping bins to easier follow the character of the physical-parameter changes. The results, as the functions of Blazhko phase, are shown in Fig.~\ref{Fig:IPM} along with photometric variations derived directly form the light curve, without the need of the IPM \citep[see details in][]{jurcsik2009c}. 

The left-hand column of Fig. \ref{Fig:IPM} shows the parameters directly derived from the multi-colour light curves, from top to bottom: the $V$ amplitude modulation in $f_0$; a rather weak but detectable phase modulation of $f_0$; variations in magnitude and intensity averaged mean $V$ brightness; variations in different magnitude and intensity averaged $B-V$ colours; and the same in $V-I$ (all averages correspond to pulsation-phase). These diagrams reveal a very weak phase modulation corresponding to $\approx 0.01$\,d variations in $O-C$. The same result was obtained from the Fourier model of the light curve in Sect. \ref{Sect:Photometry}. We also can observe from Fig. \ref{Fig:IPM} that only the intensity-averaged magnitudes and their differences show variations with the modulation. Namely, that Z~CVn is the brightest in $V$ and the warmest at high-amplitude Blazhko phase. This is a general property of all the Blazhko stars with similar detailed analyses (see the previous IPM results listed at the beginning of this section).

The right-hand panels of Fig. \ref{Fig:IPM} show the results obtained by the IPM, from top to bottom: mean radius, mean $\log g$, magnitude and intensity-averaged mean $M_{V}$, mean luminosity and mean effective surface temperature (all averages correspond to pulsation-phase). Since the distance is obviously Blazhko-phase independent, $M_{V}$ variations directly follow from $V$ variations in the right-hand column. Apart from $M_{V}$, only the luminosity show clear variations with \bl\ phase, and similarly to all previously investigated Blazhko stars, Z~CVn is the most luminous at high-amplitude \bl\ phase. The other parameters do not show significant variations with modulation phase compared to their uncertainty, which is estimated from the spread of the data points obtained with four different IPM settings. Temperature variation is not evident, but the data hints a marginally higher temperature at high-amplitude \bl\ phase in accordance with the colour variations shown in the two bottom right-hand panels.

Comparing Fig. \ref{Fig:IPM} with fig. 11 in \cite{jurcsik2012b}, we find remarkable agreement between the modulation behaviour of Z CVn and RZ~Lyrae. Each parameter changes -- or remains constant -- very similarly. The only difference is that the changes of Z CVn are roughly half of those of RZ~Lyr in each parameter. Z~CVn is similarly metal poor as RZ~Lyr, but its pulsation period is much longer than that of RZ~Lyr (0.654\,d vs. 0.511\,d).

The IPM results slightly differ from the results based on the empirical calibrations (see Tab. \ref{Tab:PhysParam}). For example, the difference in $M_{V}$ is only 0.057\,mag, which does not exceed 2 sigma. The most probable explanation of this discrepancy is that parameters calculated from empirical formulae with relatively large uncertainties were used as input parameters of the IPM. In addition, IPM works directly with the light curves in different passbands, thus any uncertainty in colours brings additional imprecision. 

\renewcommand{\arraystretch}{1.2}
\setlength\tabcolsep{3.5pt}
\begin{table*}
		\centering
		\caption{The mean physical parameters of Z CVn. BL-free in the first column means light-curve fit after the modulation was subtracted, and IPM means inverse photometric method. The errors of IPM values come from four different settings \citep[see details in][]{sodor2009a} and are certainly underestimated.}		
	\begin{tabular}{ccccccccc}\hline 
&	[Fe/H]$_{\rm JK}$\,[dex]	&	log $L$\,[L$_{\odot}$]	&	$M_{V}$\,[mag]	&	$(B-V)_{0}$\,[mag]	&	log $g$\,[dex]	&	log $T_{\rm eff}$	&	$\mathfrak{M}$\,[$\mathfrak{M}_{\odot}$]	&	$d$ [pc]	\\ \hline
BL-free &	-1.64(12)	&	1.638(12)	&	0.54(3)	&	0.351(1)	&	2.699(1)	&	3.805(2)	&	0.59(16)	&	1760(120)	\\
IPM		& -1.64, fixed &	1.753(8)	&	0.597(1)	&		&	2.6229(5)	&	3.8009(2)	&	0.59, fixed	& 1760, fixed\\ \hline
	\end{tabular}\label{Tab:PhysParam}
	\end{table*}

\section{Discussion of the binary hypothesis}\label{Sect:BinaryHypothesis}

Several RRL binary candidates are proposed to have companions with masses in a black-hole range \citep{derekas2004,liska2016b,sodor2017}, although all the respective authors considered this possibility with strong scepticism. However, the minimal mass of Z CVn companion calculated from our LTTE model (57\,$\mathfrak{M}_{\odot}$) seems to be very unlike or unrealistic even in comparison with the other candidates. Before the detection of the gravitational waves produced by the coalescence of two black holes, the highest observed mass of a BH was about 15\,$\mathfrak{M}_{\odot}$ \citep[member of X-ray binary M33 X-7,][]{ozel2010}. \citet{abbott2016} showed that the event GW150914 detected by the Laser Interferometric Gravitational-Wave Observatory (LIGO) was a product of merging of two BH with masses of 36 and 29\,$\mathfrak{M}_{\odot}$. This is a clear evidence that also `heavy' stellar BHs must exist and that there must be a way to produce such massive BHs.

Theoretical studies by \citet{spera2015} and \citet{belczynski2016} showed that at low metallicity ($Z<1/30$\,$Z_{\odot}$) the mass loss in very massive stars with mass larger than 25\,$\mathfrak{M}_{\odot}$ is strongly reduced giving enough space to form BHs with masses in the order of tens of solar masses. In addition, if strong magnetic field is present, the mass loss driven by the strong stellar wind is even more effectively reduced, thus, further increasing the chance to form `heavy' BHs \citep{petit2017}. These studies show that progenitors with sufficient mass that could form a BH with 57\,$\mathfrak{M}_{\odot}$ are theoretically possible. Because the metallicity of the companion is assumed the same as for Z~CVn ($Z\sim0.0003$, see sect. \ref{Sect:PhysChar}), creation of heavy BH cannot be rejected.

The question now is whether we can detect the BH companion of Z CVn. If there were sufficient amount of material falling onto the BH, it would produce detectable high-energy radiation. There are three possible sources of matter in the hypothetical Z CVn-BH system. If the system were close enough, the RRL component could feed the BH via classical mass transfer through the inner Lagrangian point -- a process well known in cataclysmic variables. But this is certainly not our case since the system is very wide. The remaining two sources of material, the stellar wind from RRL and the surrounding interstellar matter (ISM), can sum up and let the BH radiate via spherical, so called Bondi-Hoyle-Littleton (BHL) accretion \citep{hoyle1939,bondi1944,bondi1952}.  

It is generally assumed that RRLs loose mass during the evolution on the horizontal branch, but the rate of the mass loss is unknown. \citet{koopmann1994} estimated the maximal mass-loss rate that would not have significant effects on the horizontal branch morphology as $10^{-9}$\,$\mathfrak{M}_{\odot}$/year. If this was the case, the BH would be orders of magnitude brighter than the RRL itself. However, we do not know the true mass-loss rate, speed of the stellar wind particles, and we do not know the speed of sound in the out-flowing material, which is defined by the density and temperature of the material. Speed of sound ($v_{\rm s}$) and relative speed of the BH and the ambient material $v_{\rm r}$ are the crucial quantities when calculating the rate of the accretion because $\dot{M}\sim (v^{2}_{\rm s}+v^{2}_{\rm r})^{-3/2}$. Thus the luminosity of the BH could be several orders of magnitude larger, but also several orders of magnitude lower than the luminosity of Z CVn.

The problem is the same with pure ISM accretion. We do not know exactly what is the density of the ISM, its temperature and whether Z CVn moves through the ISM or co-rotates with it around the Galactic center ($v_{\rm r}=0$). If we assume the density $n=0.02$ cm$^{-3}$ \citep[fig. 1 in][for $Z=1.8$\,kpc \citep{maintz2005}]{cox2005} the resulting luminosity ranges from 10$^{-5}$ to $10^{4}$\,L$_{\odot}$ depending on the temperature ($10^{4}$--$10^{5}$\,K) and $v_{\rm r}=0$ or $v_{\rm r}=178$\,km/s \citep[calculated from the galactocentric velocities,][]{maintz2005}. Obviously, the accretion rate cannot be calculated properly. In any case, in the direction to Z CVn no X-ray or gamma ray source has been identified, thus the confirmation of the presence of BH fails anyway.

If the star exploded as a supernovae one can search for the envelope ejected during the explosion. This idea also faces serious problems. Very massive stars ($>40$\,$\mathfrak{M}_{\odot}$) probably collapse directly to BH without exploding as supernovae \citep{fryer1999}. Even if the star exploded, it necessarily had to occur very shortly after formation of the system, i.e. billions of years ago. Since lifetime of the supernovae remnants is certainly less than this time, there is no chance to detect it at present days. Therefore, the chance to detect the supernova remnant is extremely low, even impossible.

We can also investigate the probability that an RRL star is bound with a 57-$\mathfrak{M}_{\odot}$ BH in wide system. Using the \citet{salpeter1955} IMF function
\begin{equation}
\xi(\mathfrak{M})=\xi_{0}\mathfrak{M}^{-2.35}
\end{equation}
we can estimate the ratio of the number of stars with $>60$\,$\mathfrak{M}_{\odot}$ (BH progenitor) and number of RRL progenitors with $\sim 1$ solar mass. This ratio is $\xi({\rm BH})/\xi({\rm RRL})\sim 6.6\cdot 10^{-5}$, thus one $60$-$\mathfrak{M}_{\odot}$ star may exist for each $\sim$15\,000 1-Solar mass stars. Not all stars are bound in binary systems. In case of RRLs it is probably less than 4\,\% \citep{hajdu2015}. With this estimate the ratio is 1:375\,000. At this point, the situation could look optimistic, because almost 100\,000 RRL stars have been identified so far, which gives good chance to detect a system with 60-solar mass BH. However, independent and unambiguous confirmation of such system would be very difficult. We have to point out here that among 30 recently identified RRL binary candidates \citep[see the online list by][]{liska2016c} with orbital periods larger than 2\,000\,d, four systems are suspected to have a companion with mass in the BH range.

\begin{figure}
	\includegraphics[width=\columnwidth]{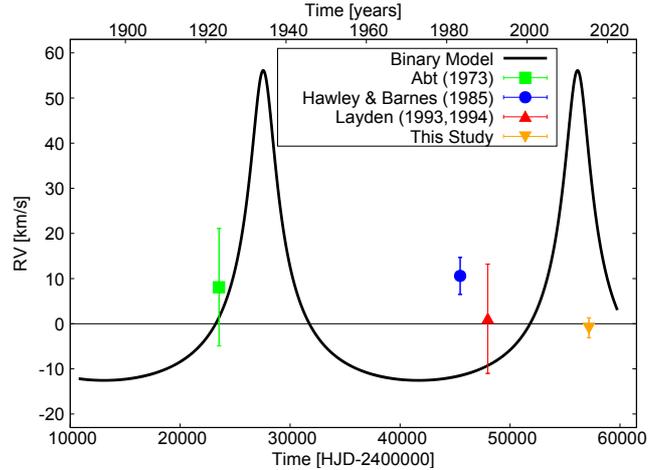}
    \caption{Synthetic $\gamma$-velocity curve (the solid black line) with the systemic velocities from Table \ref{Tab:RVgvelTable}. No offset corresponding to the velocity of binary centre-of-mass is added.}
    \label{Fig:RVOrbit}
\end{figure}

All the above discussion is only very uncertain, but we have strong evidence why to reject the RRL-BH hypothesis. From the model of the proposed orbit (parameters are in Table \ref{Tab:LiTEparam}) we calculated the systemic RV curve and compared it with the RV values determined from the available measurements (Table \ref{Tab:RVgvelTable}). The result is shown in Fig. \ref{Fig:RVOrbit}. Obviously, the observations do not match the theoretical RV curve at all. Our measurements from 2015-2016 should be shifted of about 50\,km/s against the older measurements to fit the model. Also RV observations from \citet{hawley1985} and \citet{layden1993,layden1994} should have different values to fit the model. 

We can safely declare that these shifts cannot be caused by the differences in usage of metallic/hydrogen lines in case when it was ambiguous \citep{abt1973,layden1993,layden1994} because this difference could be maximally about 5\,km/s \citep[estimated from different RV templates by][]{sesar2012}. The other source of systematic errors could be the presence of the \bl~effect. According to our new observations that were obtained in different \bl~phases, the scatter in RVs due to the \bl~effect cannot be more than about 10\,km/s. Therefore, the proposed binary hypothesis can be safely rejected, because even if we assume the highest possible systematic errors, the observations cannot fulfil the model expectations.  

\begin{figure}
	\includegraphics[width=\columnwidth]{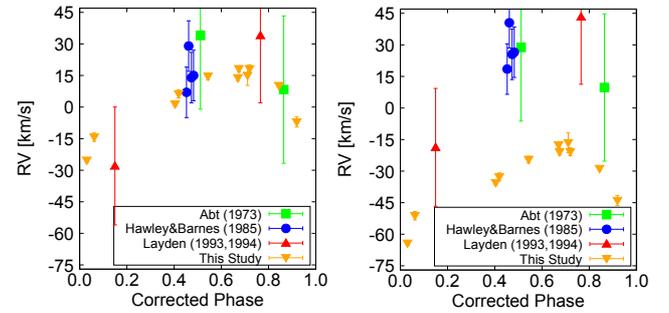}
	\caption{RV-observations folded with the pulsation period that was corrected from the LTTE using our \oc~model. When also the model of the $\gamma$-velocity is removed, the points should stitch together. The opposite is observed (the right-hand panel).}
    \label{Fig:RV_phases}
\end{figure}

The same is seen from the Fig. \ref{Fig:RV_phases}, where the RV observations are plotted against time-corrected phase (LTTE subtracted). After the RV observations are corrected for the orbital motion (the right-hand panel), the scatter should decrease. The opposite is seen from that figure. These two facts are strong evidences that the binary hypothesis is false and that the presumed cyclic long-term period variation is not caused by the LTTE, but by some different physical process that is intrinsic to the star. In fact, our results are consistent with a single star of 0\,km/s centre-of-mass RV.

\section{Summary and conclusions}\label{Sect:Discussion}

We present a comprehensive study of the period evolution of Z CVn and its light curve using original photometric and spectroscopic measurements, as well as observations from literature. The star was observed in 53 nights photometrically (four seasons) and in 6 night spectroscopically (two seasons). We caught 18 maximum times and additional 209 were taken from the literature and from the GEOS database creating a set of 227 maximum times (Tab. \ref{Tab:MaximumTimes}) covering more than one century.

The shape of the general trend in the \oc~diagram suggests that LTTE could be present. The detailed model of the period variations with the assumption of LTTE (Table \ref{Tab:LiTEparam}) shows that the companion should be a BH with a mass of about $\mathfrak{M}_{\rm 2, min} \!\sim\! 57$\,$\mathfrak{M}_{\odot}$. Although such RRL-BH configuration is theoretically possible, we do not have any clear evidence that this is the case of Z CVn. There is also no observational evidence of the BH (X-ray radiation) in the direction to Z~CVn.

However, the rejection of the binary hypothesis was firmly confirmed by the spectroscopic measurements, which do not match the theoretical $\gamma$ velocity at all. A further test with time-corrected phases and RV observations confirms this finding too. Thus, our analysis gives firm proofs that possibly cyclic variations of the pulsation period of Z CVn cannot be explained as LTTE caused by an unseen companion. This finding has large impact on the other known binary candidates with similar periods and period-variation-amplitudes, because it shows that there may be some unknown mechanism that causes large-amplitude period variations. Already \citet{firmanyuk1982} pointed out that such variations are probably not a result of LTTE. Perhaps some analogue to the \bl~effect could be present, but with much longer modulation period. We do not have any physical explanation for such long period variations and formerly proposed scenarios for the long-term period variations are also unlikely (Sect.~\ref{Sect:PeriodEvolution}).

On the basis of our photometric observations we estimated the basic pulsation period, but also detected the \bl~effect with period of 22.391\,d. This period was clearly detected in maximum brightness variation. Peaks suggesting additional modulation with period comparable with the length of the data set were also detected, but they could only be artefacts of the secular period change. We found that the modulation is almost exclusively in amplitude (with amplitude of about 0.3\,mag) with period/phase modulation less than 0.01\,d. This is in contrast with previous studies, when also period/phase modulation was detectable \citep{kanyo1966,leborgne2012}. Because the pulsation period is comparable with the one from \citet{kanyo1966}, but the period/phase modulation is undetectable and the \bl~period is longer than at Kany\'{o}'s time, our study contradicts the anti-correlation between the length of \ppuls~and \pbl~proposed by \citet{leborgne2012}. However, the fact that \pbl~continuously changes together with the major period variations means that these two effects might be somehow connected. This is another evidence against the LTTE assumption because possible companion at the orbit with 80\,year-long period certainly could not tidally influence the pulsation properties of the RRL component.     

Although our spectroscopic measurements are of insufficient quality and coverage for reliable physical parameter determination, we estimated them from the shape of the light curve. The values based on modulation-free light curve are in excellent agreement with formerly determined values in literature \citep[our ${\rm [Fe/H}=-1.64$ vs. $-1.7$\footnote{After recalculating from \citet{layden1994} to \citet{jurcsik1996} metallicity scale using the formula from \citep{sandage2004}.}, our $r=1760$\,pc vs. 1770\,pc,][]{layden1994}. This experience suggests that parameters estimated from modulation-free light curve are of better reliability than those based on the mean-light curve fitting or averaging parameters over the Blazhko cycle.

The star now lengthens its pulsation period and between HJD 2460000 and 2462500 is predicted to reach its maximum (if the period variation is really cyclic). After that it should shorten again. Monitoring of Z CVn is, therefore, desired and should prove the stability of the long-period cycle in Z CVn. If the long cycle is really intrinsic to the star, new theoretical explanation will be needed to explain not only behaviour of Z CVn, but also of other RRL stars \citep[e.g. some candidates from][]{liska2016b,sodor2017}.

\section*{Acknowledgements}
We are pleased to express our gratitude to Ond\v{r}ej Pejcha, Lenka Zychov\'a and Norbert Werner for the discussion about black holes, Bondi-Hoyle-Littleton accretion and interstellar matter. MS acknowledges the support of the postdoctoral fellowship programme of the Hungarian Academy of Sciences at the Konkoly Observatory as host institution. The financial support of the Hungarian NKFIH Grants K-115709, K-113117 and K-119517 is acknowledged. \'{A}S was supported by the J\'{a}nos Bolyai Research Scholarship of the Hungarian Academy of Sciences. This research was carried out under the project CEITEC 2020 (LQ1601) with financial support from the Ministry of Education, Youth and Sports of the Czech Republic under the National Sustainability Programme II (JL). Travel support from McDonald Observatory to obtain the spectroscopy observations is gratefully acknowledged (TGB). The support of Brno Observatory and Planetarium is acknowledged. The research made use of the International Variable Star Index (VSX) database, operated at AAVSO, Cambridge, Massachusetts, USA, NASA's Astrophysics Data System. We thank J.-F. LeBorgne and other people from GEOS for maintaining their RRL database. 

\bibliographystyle{mnras}
\bibliography{references}

\appendix

\section{Physical parameters determination based on the light-curve shape}\label{Sect:PhysChar_App}

First we divided the data according to their \bl~phase into ten segments (blue points in subplots in Fig. \ref{Fig:BLPhased}, using $P_{\rm BL}=22.931$\,d and the zero epoch from eq. \ref{Eq:PulsElements}). Each subset was fitted using eq. \ref{Eq:LCModel} with ten pulsation components (harmonics of the inverse mean period from eq. \ref{Eq:PulsElements}) to get the mean light curve during each of the \bl~phases shown with the full red line in Fig. \ref{Fig:BLPhased}. The used pulsation-phase-independent Fourier parameters follow the classical definition by \citet{simon1981} and are computed as
\begin{equation}
	R_{i1}=A_{i}/A_{1} ~~ {\rm and} ~~ \phi_{i1}=\phi_{i}-i\phi_{1},
\end{equation}
where $A_{i}$ and $\phi_{i}$ are amplitudes and phases of particular frequency harmonic. Resulting Fourier parameters for each \bl~phase can be found in Table \ref{Tab:PhysChar_App} together with physical parameters that were computed using empirical formulas taken from the following papers:

Metallicity \citep{jurcsik1996}:
\begin{equation}
	{\rm [Fe/H]}_{\rm JK}=-5.038-5.394P_{\rm Puls}+1.345\phi_{31},
\end{equation}

Luminosity \citep{jurcsik1998}:
\begin{equation}
	\log L=1.464-0.106{\rm [Fe/H]}_{\rm JK},
\end{equation}

Absolute magnitude \citep{catelan2008}:
\begin{equation}
	M_{V}=0.23{\rm [Fe/H]}_{\rm ZW}+0.984,
\end{equation}

Extinction-free colour index \citep{jurcsik1998}:
\begin{equation}
	(B-V)_{0}=0.308+0.163P_{\rm Puls}-0.187A_{1},
\end{equation}

Surface gravity \citep{jurcsik1998}:
\begin{equation}
	\log g=2.473-1.226\log P_{\rm Puls}
\end{equation}

Effective temperature \citep{kovacs2001}:
\begin{equation}
	\log T_{\rm eff}=3.884-0.3219(B-V)_{0}+0.0167\log g + 0.007{\rm [Fe/H]}_{\rm JK},
\end{equation} 

Mass \citep{jurcsik1998}:
\begin{equation}
	\log \mathfrak{M}=-0.328-0.062{\rm [Fe/H]}_{\rm JK}.
\end{equation}

Metallicity scale  ${\rm [Fe/H]}_{\rm JK}$ in these relations is from \citet{jurcsik1998}, while [Fe/H]$_{\rm ZW}$ is from \citet{zinn1984}. The relation between these two scales is
\begin{equation}
	{\rm [Fe/H]}_{\rm ZW}=1.05{\rm [Fe/H]}_{\rm JK}-0.2,
\end{equation}
\citep{sandage2004}. 

We can also estimate the distance to Z CVn. 

The parameters cannot be calculated in \bl~phase 0.7 because of poor coverage. Also phase 0.4 is badly sampled and the parameters significantly differ from those in the other \bl~phases (see Fig. \ref{Fig:BLPhased} and Table \ref{Tab:PhysChar_App}). Therefore, values coming from these data were omitted when calculating the average values.

The distance was simply calculated using
\begin{equation}
	\log d=\frac{V-M_{V}-A_{V}+5}{5},
\end{equation}
where $M_{V}$ is the visual absolute magnitude, $V$ is the observed mean magnitude and $A_{V}$ is the interstellar extinction at the center of $V$ filter. Assuming $M_{V}=0.54$\,mag, $V=11.82$\,mag, and $A_{V}=0.05$\,mag \citep{schlegel1998,schlafly2011}, the distance is 1760(120) pc.

\renewcommand{\arraystretch}{1.2}
\setlength\tabcolsep{3.5pt}
\begin{table*}
		\centering
		\caption{The first two amplitude and phase Fourier parameters together with calculated physical characteristics. Obviously, metallicity and mass cannot change during \bl\ cycle, they are given only for completeness.}		
	\begin{tabular}{MMMMMMMMMMMMM}\hline 
Blazhko phase	&	$R_{21}$	&	$R_{31}$	&	$\phi_{21}$\,[rad]	&	$\phi_{31}$\,[rad]	&	[Fe/H]\,[dex]	&	log $L$\,[L$_{\odot}$]	&	$M_{V}$\,[mag]	&	$(B-V)_{0}$\,[mag]	&	log $g$\,[dex]	&	log $T_{\rm eff}$	&	$\mathfrak{M}$\,[$\mathfrak{M}_{\odot}$]	\\ \hline
0.0	&	0.473	&	0.301	&	2.395	&	5.053	&	-1.769	&	1.651	&	0.511	&	0.339	&	2.699	&	3.808	&	0.605	\\
0.1	&	0.474	&	0.299	&	2.452	&	5.102	&	-1.703	&	1.645	&	0.527	&	0.343	&	2.699	&	3.807	&	0.599	\\
0.2	&	0.474	&	0.242	&	2.534	&	5.334	&	-1.391	&	1.611	&	0.602	&	0.353	&	2.699	&	3.806	&	0.573	\\
0.3	&	0.444	&	0.224	&	2.509	&	5.278	&	-1.466	&	1.619	&	0.584	&	0.357	&	2.699	&	3.804	&	0.579	\\
0.4	&	0.440	&	0.232	&	2.569	&	5.662	&	-0.950	&	1.565	&	0.709	&	0.361	&	2.699	&	3.806	&	0.538	\\
0.5	&	0.421	&	0.238	&	2.470	&	5.406	&	-1.294	&	1.601	&	0.625	&	0.363	&	2.699	&	3.803	&	0.565	\\
0.6	&	0.496	&	0.264	&	2.436	&	5.348	&	-1.373	&	1.610	&	0.606	&	0.364	&	2.699	&	3.803	&	0.572	\\
0.7	&	-	&	-	&	-	&	-	&	-	&	-	&	-	&	-	&	-	&	-	&	-	\\
0.8	&	0.483	&	0.309	&	2.410	&	5.039	&	-1.788	&	1.653	&	0.506	&	0.347	&	2.699	&	3.805	&	0.606	\\
0.9	&	0.479	&	0.291	&	2.396	&	5.080	&	-1.732	&	1.648	&	0.520	&	0.340	&	2.699	&	3.808	&	0.602	\\ \hline
Average	&	{\tiny 0.47(2)}	&	{\tiny 0.27(3)}	&	{\tiny 2.45(5)}	 &	{\tiny 5.21(12)} 	&	{\tiny -1.56(20)}	&	{\tiny 1.63(2)}	&	{\tiny 0.56(5)}	&	{\tiny 0.351(10)}	&	{\tiny 2.699(1)	} &	{\tiny 3.805(2)} 	&	{\tiny 0.59(2)}	\\ \hline

Mean LC & {\tiny 0.463(5)} & {\tiny 0.255(5)} & {\tiny 2.46(2)} & {\tiny 5.21(2)} & {\tiny -1.56(6)} & {\tiny 1.629(7)} & {\tiny 0.56(2)} & {\tiny 0.351(1)} & {\tiny 2.699(1)} & {\tiny 3.805(1)} & {\tiny 0.59(8) } \\ \hline

BL-free & {\tiny 0.447(2)} & {\tiny 0.269(2)} & {\tiny 2.42(7)} & {\tiny 5.15(9)} & {\tiny -1.64(12)} & {\tiny 1.638(12)} & {\tiny 0.54(3)} & {\tiny 0.351(1)} & {\tiny 2.699(1)} & {\tiny 3.805(1)} & {\tiny 0.59(2) } \\ \hline


\hline
	\end{tabular}\label{Tab:PhysChar_App}
	\end{table*}

\section{Full light-curve solution and used maximum times}

\newcolumntype{M}{>{\footnotesize}c}
\renewcommand{\arraystretch}{1.2}
\setlength\tabcolsep{3.5pt}
\begin{table*}
		\centering
		\caption{The full solution of the light curves in different passbands. The values come from the best fit to the data.}		
	\begin{tabular}{MMMMMMMMMMMMM}\hline 

ID &	\multicolumn{3}{c}{{\it B}}&						\multicolumn{3}{c}{{\it V}}&						\multicolumn{3}{c}{{\it R}}&						\multicolumn{3}{c}{{\it I}}\\					
&	$f$\,[c/d]	&	$A$\,[mag]	&	$\phi$\,[rad]	&	$f$\,[c/d]	&	$A$\,[mag]	&	$\phi$\,[rad]	&	$f$\,[c/d]	&	$A$\,[mag]	&	$\phi$\,[rad]	&	$f$\,[c/d]	&	$A$\,[mag]	&	$\phi$\,[rad]	\\ \hline
$f_{0}-f_{\rm BL}$ &	1.485565	&	0.0452	&	3.7130	&	1.485552	&	0.0359	&	3.7189	&	1.485565	&	0.0291	&	3.7498	&	1.485561	&	0.0205	&	3.8117	\\
$f_{0}-f_{\rm M}$ &	1.528252	&	0.0231	&	1.9522	&	1.528267	&	0.0210	&	1.9723	&	1.528252	&	0.0147	&	2.0658	&		&		&		\\
$f_{0}$ &	1.529170	&	0.4448	&	4.0907	&	1.529169	&	0.3407	&	4.0506	&	1.529169	&	0.2688	&	3.9679	&	1.529169	&	0.2128	&	3.8179	\\
$f_{0}+f_{\rm BL}$ &	1.572775	&	0.0370	&	3.9215	&	1.572786	&	0.0312	&	3.9107	&	1.572773	&	0.0246	&	3.9614	&	1.572776	&	0.0186	&	3.9812	\\
$2f_{0}-f_{\rm BL}$ &	3.014735	&	0.0263	&	3.5113	&	3.014721	&	0.0208	&	3.4969	&	3.014735	&	0.0173	&	3.5151	&	3.014730	&	0.0128	&	3.4917	\\
$2f_{0}-f_{\rm M}$ &	3.057422	&	0.0195	&	2.1554	&	3.057436	&	0.0126	&	2.7507	&	3.057421	&	0.0103	&	2.2792	&		&		&		\\
$2f_{0}$ &	3.058340	&	0.2022	&	4.2486	&	3.058338	&	0.1525	&	4.2404	&	3.058339	&	0.1264	&	4.2352	&	3.058338	&	0.1005	&	4.1627	\\
$2f_{0}+f_{\rm BL}$ &	3.101945	&	0.0286	&	3.9690	&	3.101955	&	0.0247	&	3.9659	&	3.101943	&	0.0183	&	3.9825	&	3.101945	&	0.0136	&	4.0445	\\
$3f_{0}-f_{\rm BL}$ &	4.543905	&	0.0295	&	3.6360	&	4.543891	&	0.0231	&	3.7130	&	4.543904	&	0.0188	&	3.7161	&	4.543899	&	0.0141	&	3.7504	\\
$3f_{0}-f_{\rm M}$ &	4.586592	&	0.0170	&	2.5104	&	4.586605	&	0.0127	&	2.5669	&	4.586591	&	0.0096	&	2.5709	&		&		&		\\
$3f_{0}$ &	4.587510	&	0.1200	&	4.6937	&	4.587507	&	0.0919	&	4.7330	&	4.587508	&	0.0752	&	4.7169	&	4.587506	&	0.0584	&	4.6486	\\
$3f_{0}+f_{\rm BL}$ &	4.631115	&	0.0259	&	4.2344	&	4.631124	&	0.0197	&	4.2931	&	4.631112	&	0.0171	&	4.3827	&	4.631114	&	0.0139	&	4.4915	\\
$4f_{0}-f_{\rm BL}$ &	6.073075	&	0.0176	&	4.4841	&	6.073060	&	0.0142	&	4.3331	&	6.073073	&	0.0124	&	4.4282	&	6.073067	&	0.0104	&	4.4810	\\
$4f_{0}-f_{\rm M}$ &	6.115762	&	0.0123	&	3.1709	&	6.115774	&	0.0077	&	3.1279	&	6.115760	&	0.0071	&	3.1759	&		&		&		\\
$4f_{0}$ &	6.116680	&	0.0680	&	5.1909	&	6.116676	&	0.0534	&	5.1848	&	6.116677	&	0.0436	&	5.2208	&	6.116675	&	0.0328	&	5.1517	\\
$4f_{0}+f_{\rm BL}$ &	6.160285	&	0.0229	&	5.0100	&	6.160293	&	0.0166	&	5.0167	&	6.160282	&	0.0143	&	5.0255	&	6.160283	&	0.0116	&	5.1934	\\
$5f_{0}-f_{\rm BL}$ &	7.602245	&	0.0139	&	5.0402	&	7.602229	&	0.0101	&	5.2303	&	7.602243	&	0.0092	&	5.1281	&	7.602236	&	0.0075	&	5.1680	\\
$5f_{0}-f_{\rm M}$ &		&		&		&	7.644943	&	0.0079	&	3.3327	&	7.644929	&	0.0053	&	3.3548	&		&		&		\\
$5f_{0}$ &	7.645850	&	0.0325	&	5.4723	&	7.645845	&	0.0275	&	5.7006	&	7.645847	&	0.0230	&	5.6589	&	7.645844	&	0.0165	&	5.5656	\\
$5f_{0}+f_{\rm BL}$ &	7.689455	&	0.0145	&	5.4992	&	7.689462	&	0.0124	&	5.3723	&	7.689451	&	0.0100	&	5.4530	&	7.689452	&	0.0085	&	5.6281	\\
$6f_{0}-f_{\rm BL}$ &	9.131415	&	0.0101	&	5.7017	&	9.131398	&	0.0071	&	5.7274	&	9.131412	&	0.0058	&	5.7233	&		&		&		\\
$6f_{0}$ &	9.175020	&	0.0184	&	5.6101	&	9.175014	&	0.0141	&	5.6243	&	9.175016	&	0.0124	&	5.7004	&	9.175013	&	0.0099	&	5.7340	\\
$6f_{0}+f_{\rm BL}$ &	9.218625	&	0.0106	&	5.9954	&	9.218631	&	0.0085	&	5.9367	&	9.218620	&	0.0074	&	6.0217	&	9.218620	&	0.0066	&	6.2323	\\
$7f_{0}-f_{\rm BL}$ &		&		&		&	10.660567	&	0.0062	&	0.1001	&	10.660581	&	0.0042	&	6.2283	&		&		&		\\
$7f_{0}$ &	10.704190	&	0.0125	&	5.9270	&	10.704183	&	0.0081	&	6.1613	&	10.704185	&	0.0070	&	6.1231	&	10.704181	&	0.0063	&	6.2027	\\
$7f_{0}+f_{\rm BL}$ &	10.747795	&	0.0084	&	0.2438	&	10.747800	&	0.0057	&	0.2936	&	10.747790	&	0.0046	&	0.3915	&	10.747789	&	0.0046	&	0.5625	\\
$8f_{0}-f_{\rm BL}$ &		&		&		&		&		&		&	12.189751	&	0.0029	&	0.7305	&		&		&		\\
$8f_{0}$ &	12.233360	&	0.0075	&	0.1595	&	12.233352	&	0.0062	&	6.2551	&	12.233355	&	0.0044	&	0.0253	&	12.233350	&	0.0037	&	0.0415	\\
$8f_{0}+f_{\rm BL}$ &		&		&		&	12.276969	&	0.0054	&	0.9788	&	12.276959	&	0.0040	&	0.9867	&		&		&		\\
$9f_{0}-f_{\rm BL}$ &		&		&		&		&		&		&	13.718920	&	0.0028	&	1.0452	&		&		&		\\
$9f_{0}$ &		&		&		&	13.762521	&	0.0038	&	0.5985	&	13.762524	&	0.0026	&	0.4936	&		&		&		\\
$9f_{0}+f_{\rm BL}$ &		&		&		&		&		&		&	13.806128	&	0.0020	&	1.3590	&	13.806127	&	0.0029	&	1.7063	\\
$10f_{0}-f_{\rm BL}$ &		&		&		&		&		&		&		&		&		&		&		&		\\
$10f_{0}$ &		&		&		&	15.291690	&	0.0041	&	0.8719	&	15.291693	&	0.0028	&	0.7726	&		&		&		\\
$10f_{0}+f_{\rm BL}$ &		&		&		&		&		&		&	15.335298	&	0.0020	&	1.7188	&	15.335295	&	0.0025	&	2.1285	\\

\hline
	\end{tabular}\label{Tab:Freq}
	\end{table*}

\bsp	
\label{lastpage}
\end{document}